\documentclass[UTF8,zihao=-4]{ctexart}

\usepackage[a4paper,margin=2.6cm]{geometry}
\usepackage{booktabs}
\usepackage{tabularx}
\usepackage{longtable}
\usepackage{array}
\usepackage{authblk}
\usepackage{enumitem}
\usepackage{hyperref}
\usepackage{csquotes}
\usepackage{graphicx}
\usepackage[backend=biber,style=numeric,sorting=none]{biblatex}
\usepackage[absolute,overlay]{textpos}% by XZ: for the text in the top margin..

\DeclareBibliographyAlias{standard}{misc}

\hypersetup{
  colorlinks=true,
  linkcolor=black,
  citecolor=black,
  urlcolor=blue
}

\AtBeginBibliography{\sloppy}
\DefineBibliographyStrings{english}{references = {References / 参考文献}}

\setlist[itemize]{leftmargin=2em,itemsep=0.2em}
\setlist[enumerate]{leftmargin=2.2em,itemsep=0.2em}

\newcommand{\en}[1]{\textit{#1}}
\newcommand{\zh}[1]{\textbf{#1}}
\newcommand{\missingfigure}[1]{%
  \fbox{\parbox[c][2.2in][c]{0.78\linewidth}{\centering #1}}%
}

\title{\heiti 此安全非彼安全：\\一个给中国计算机科学与工程学者的术语倡议}
\author{赵星宇\\
  \small 国家网络安全学院，武汉大学\footnote{本文仅代表作者个人观点，不代表作者所属组织、单位或相关机构的立场。}\\
  \small \texttt{xingyu.zhao@whu.edu.cn}
  }
\date{\today}

\begin{document}

\maketitle

\begin{textblock*}{20cm}(1cm,1cm)
\textcolor{black}{The English version of this Chinese paper is included from Section 10 onward.}
\end{textblock*}

\begin{abstract}
在中文计算机科学与工程语境中，{safety} 与 {security} 长期被共同译作“安全”。这一译法在日常交流中简洁，却在标准解读、跨学科协作、风险分析和论文写作中造成持续的概念折叠：当研究者需要同时讨论系统是否免于不可容忍的非蓄意危害，以及系统是否能抵御对抗性威胁时，“安全”一词往往无法承担区分责任。本文主张，在保留既有标准名称和法律术语的同时，学术写作和工程文档应将 {security} 统一译作“安保”或“安保性”，将 {safety} 主要译作“安全”或“安全性”。这一倡议并非简单的文字替换，而是面向科学认知、工程风险沟通和安全论证的术语治理方案。本文首先梳理国际与国内标准中 {safety}/{security} 的概念边界，分析“同译为安全”在功能安全、预期功能安全、信息安全、网络安全、汽车网络安全和人工智能治理中的表达损耗。并针对人工智能系统，结合国际上最新关于AI assurance、safety-security co-assurance和security-informed safety的近年研究，说明准确术语对于形成可检验、可质疑、可沟通的科学论证具有基础意义。最后，本文提出分阶段、双轨制的中文术语实践建议。
\end{abstract}

\noindent\textbf{关键词：}术语翻译；安全；安保性；AI assurance；Assurance 2.0；security-informed safety；功能安全；预期功能安全

\section{引言}

2023 年 11 月 2 日，英国政府在首届人工智能安全全球峰会（AI Safety Summit）期间宣布成立 AI Safety Institute，称其任务是评估前沿人工智能模型的风险\parencite{govuk2023AISIlaunch,govuk2023AISIoverview}。一年多以后，2025 年 2 月 14 日，英国政府宣布该机构更名为AI Security Institute，强调新机构将把国家安全、犯罪滥用和相关保护工作纳入其核心使命\parencite{govuk2025AISIRename}。AISI 缩写未变，牌子上的关键名词却从 {safety} 变成了 {security}；
%机构一周年总结中对模型评估、研究能力和公共沟通的描述，也显示这个机构一直处在 AI 风险分类和治理边界不断调整的过程中\parencite{aisi2024FirstYear}。
这个例子生动地说明了一件事：在英语科技政策和工程治理语境中，safety与security并不是可以随意互换的近义词。它们都关心“坏事不要发生”，但坏事的来源、风险建模方式、工程证据、组织责任和治理手段并不相同。但在中文科技语境中，计算机科学与工程的学者们长期把二者都译作“安全”。于是，AI Safety、AI Security、functional safety、SOTIF、cybersecurity、information security、system safety engineering、system security engineering 等都被压缩进同一中文词根：“安全”。

\begin{figure}[ht]
    \centering
    \IfFileExists{pic/aisi.png}{%
      \includegraphics[width=0.8\linewidth]{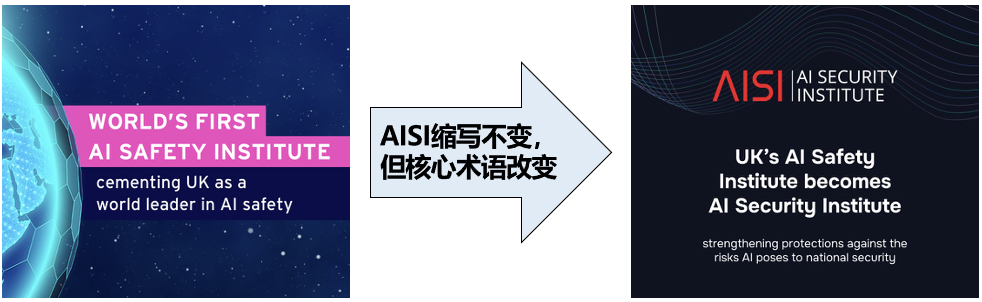}%
    }{%
      \missingfigure{Figure file \texttt{pic/aisi.png} not found. The Overleaf/arXiv version will use the original AISI image if the file is present.}%
    }
    \caption{英国AISI改名展示的概念边界：同一政府机构从safety改牌为security，说明二者在风险分类、治理边界和组织任务中承担不同概念功能。}
    \label{fig_aisi}
\end{figure}

这种压缩的代价在单一领域内也许可以靠上下文弥补；但在软件定义汽车、工业控制、医疗器械、机器人、航空航天和通用人工智能等交叉系统中，safety与security的交互日益成为核心问题。一个车辆感知系统可能没有随机硬件故障，却因性能局限造成safety风险；同一个系统也可能因远程攻击、数据投毒或权限绕过造成security风险。若二者都只叫“安全风险”，风险评审会议、需求文档、论文标题和标准索引都会变得含混。这种语义上的重叠与模糊，不仅给科研交流带来额外的沟通成本负担，也可能在关键概念理解上引发歧义，甚至误导风险判断与决策。

本文提出一个有意略显“逆耳”的倡议：在计算机科学与工程研究中，将security统一译为“安保”或“安保性”，将safety继续译为“安全”或“安全性”。“安保性”借鉴了国际关系和公共治理中“安全保卫/安保”的使用传统。例如中国驻秘鲁使馆与秘鲁国家警察、秘鲁中资企业协会建立的“中秘双边安保联席会议机制”\parencite{mfa2022ChinaPeruAnbaoMechanism}，这里的“安保”显然指向围绕人员、机构和项目保护的 {security} 措施。“安保”比“安全”更突出保护、防护、守卫和对抗性威胁这一层含义。本文并不主张立即改写法律、国家标准和已经固定的行业名称，而是主张在未来的计算机科学与工程领域的学术论文、教材、术语表和工程文档中，用“安保性”为 security确立一个稳定的中文位置。

\section{术语与认知}

术语问题容易被误解为“叫法”问题。但在科学研究中，术语并不是贴在已有对象上的标签，而是研究共同体识别对象、划分边界、组织证据和分配责任的工具。维特根斯坦在《逻辑哲学论》中提出“语言的界限意味着世界的界限”\parencite{wittgenstein1922Tractatus}；语言相对论的温和版本也并不要求接受“语言决定思想”的强命题，而是指出语言中的词汇和范畴会成为概念化与行动的惯性路径\parencite{whorf1956Language,lucy1998SapirWhorf}。换言之，人当然可以在没有某个词的情况下遭遇某种经验，但若一个共同体长期缺少稳定词项，它就更难把该经验变成可检索、可教学、可审稿、可标准化的研究对象。

这正是safety/security同译为“安全”的深层问题。中文研究者并非不知道二者有差异；真正的问题是，公共词汇没有给这种差异一个低成本、可持续的表达位置。于是，差异每次都必须依赖上下文临时解释，不能稳定沉淀为标题、关键词、课程模块、标准索引、需求字段和评审清单。Bowker 与 Star 对分类系统的研究提醒我们，分类一旦进入工作流和基础设施，就会反过来塑造组织能看见什么、忽略什么、记录什么\parencite{bowkerStar1999SortingThingsOut}。在工程领域，术语错误不是修辞瑕疵，而可能变成风险登记、责任归属和证据结构的错误。

因此，本文所倡议的“安保性”并不是为了制造新词，而是为了给一个已经存在、且越来越重要的概念族提供中文认知支点。若没有这个支点，国内外现有的和最先进的研究方向如{safety case} 与{security case}、{AI safety}与{AI security}、{safety-security co-assurance}、{security-informed safety} 等等都会在中文里反复坍缩成“安全论证”“人工智能安全”“安全-安全协同保证”“安全知悉的安全”。这种坍缩会限制研究者提出问题的方式：如果语言总是把两个对象合成一个对象，研究者就必须先花额外成本把它们重新拆开。

\section{概念边界}

正因为分类系统会塑造工程师能“看见”什么，safety与security的范畴划定才不是事后命名，而会直接影响危害分析、威胁建模和论证边界。若二者在中文中被同一个“安全”吸收，工程师仍可凭经验临时区分，但文档、标准、工具链和评审程序却很难稳定记录这种区分。下面的标准谱系梳理，正是为了说明这种概念差异在工程实践中已经对应着不同的问题结构、证据类型和责任分工。

从可信性“dependability”的经典分类看，safety与security本来也不是同一个属性的两个说法，而是可靠性reliability、可用性availability、完整性integrity、可维护性maintainability等属性网络中的相邻概念\parencite{avizienis2004Dependable}。它们相邻，所以需要协同；它们不同，所以不能被同一个中文词无差别吸收。

\subsection{Safety：系统不要伤害人和资产}

直观地说，{safety} 关心的是：\textbf{系统在其预期使用和可预见误用条件下，不应对人、财产、环境或任务造成不可容忍的伤害}。它关注的典型问题不是“有没有攻击者”，而是系统自身的失效、性能不足、设计缺陷、环境扰动、人机交互错误或组织过程失灵，如何把一个系统带入危险状态。换言之，{safety} 首先问的是：\textbf{系统会不会伤害它所处的环境？}

因此，{safety} 的风险源通常是危害源、失效模式、功能不足、操作错误、可预见误用和复杂系统交互，而不是一个有意志、有目标、会学习的攻击者。它的工程方法也相应围绕危害分析、风险降低、安全完整性、验证确认和安全论证展开。ISO/IEC Guide 51 将 {safety} 定义为免于不可容忍的风险，并把风险理解为伤害发生概率与伤害严重程度的组合\parencite{isoGuide51,kishimoto2015Guide51}。系统安全研究由此发展出对事故因果、组织过程和控制结构的系统性分析\parencite{leveson2012SaferWorld}。

在工程领域，这一路线形成了庞大的标准谱系。IEC 61508 处理电气/电子/可编程电子安全相关系统的功能安全（{functional safety}），其中软件安全要求也是独立的重要部分\parencite{iec61508-1,iec61508-3}；国内 GB/T 20438 则对应这一功能安全标准族\parencite{gbt20438-1}。ISO 26262 将功能安全扩展到道路车辆 E/E 系统，国内 GB/T 34590 也沿用了“功能安全”的中文名称\parencite{iso26262-1,gbt34590-1}；ISO 21448 和国内 GB/T 43267 则处理“预期功能安全”（SOTIF），即没有系统故障时由功能不足、性能局限或可预见误用导致的不合理风险\parencite{iso21448,gbt43267}。这些标准的共同主题不是“保密”或“防入侵”，而是通过危害分析、风险降低、安全完整性、验证确认和安全论证，使系统达到可容忍风险。

\subsection{Security：系统不要被环境中的威胁破坏}

与此相对，{security} 关心的是：\textbf{系统、数据、资产、服务和任务不应被外部或内部的威胁、攻击、滥用、未授权访问或对抗性行为破坏}。若说 {safety} 首先问“系统会不会伤害环境”，那么 {security} 首先问的是：\textbf{环境中的威胁会不会伤害系统、资产和任务？} 本文以下将 {security} 统一译为“安保”或“安保性”。

因此，安保性关注的核心不是随机失效或性能不足本身，而是资产、威胁主体、攻击面、脆弱性、权限、认证、访问控制、供应链、数据完整性和服务可用性。ISO/IEC 27000 严格来说定义的是 {information security}，即保持信息的保密性、完整性和可用性，并指出真实性、问责性、不可否认性和可靠性等属性也可能包含其中；ISO/IEC 27001 则把这些目标制度化为信息安全管理体系要求\parencite{isoIec27000,isoIec27001}。Saltzer 和 Schroeder 关于信息保护的经典论文讨论的是如何防止未授权使用或修改\parencite{saltzer1975Protection}；经典安保工程教材也围绕威胁、访问控制、协议、软件漏洞和系统防护展开\parencite{anderson2020SecurityEngineering,bishop2018ComputerSecurity}。NIST Cybersecurity Framework 2.0 面向组织管理网络安保风险\parencite{nistCSF2024}；NIST SP 800-160 则从系统工程和网络韧性角度处理可信安保系统的安保工程问题\parencite{nistSP800160v1r1,nistSP800160v2r1}。

安保分析也形成了与危害分析不同的方法族，例如误用例和攻击树分别把需求缺口与攻击路径显式化\parencite{sindre2005MisuseCases,schneier1999AttackTrees}。在行业标准中，ISO/SAE 21434 针对道路车辆网络安保工程，要求在概念、开发、生产、运行、维护和退役全生命周期中管理网络安全风险\parencite{isoSae21434}；UNECE R155/R156 和 ISO 24089 则进一步把车辆网络安保管理、软件更新和更新管理体系制度化\parencite{unR155,unR156,iso24089}；IEC 62443 系列针对工业自动化与控制系统的安保，覆盖资产所有者、系统集成商和产品供应商\parencite{iec62443-1-1,iec62443-2-1}。国内 GB/T 25069《信息安全技术 术语》、GB/T 22239《网络安全等级保护基本要求》、GB/T 35273《个人信息安全规范》以及汽车网关、车载信息交互、电动汽车远程服务系统的信息安保标准，均落在这一大类\parencite{gbt25069,gbt22239,gbt35273,gbt40855,gbt40856,gbt40857}。表\ref{table_summary_key_diff}从多个维度总结了两概念之间的差异，同时表\ref{table_standards}总结了常见的安全与安保标准和其关注范围。

\begin{table}[htbp]
\centering
\caption{{safety} 与 {security} 的典型工程差异}
\begin{tabularx}{\textwidth}{>{\raggedright\arraybackslash}p{0.18\textwidth}XX}
\toprule
维度 & Safety：安全性 & Security：安保性 \\
\midrule
主要问题 & 系统如何避免造成不可容忍的伤害 & 系统如何保护资产免受威胁、滥用和未授权行为 \\
风险来源 & 危害源、失效、性能局限、环境扰动、可预见误用、组织缺陷 & 攻击者、滥用者、内部人、恶意软件、供应链、未授权访问、对抗样本 \\
分析对象 & 危害、危险状态、事故场景、风险降低措施 & 资产、威胁、脆弱性、攻击面、控制措施 \\
典型方法 & HARA、FMEA、FTA、STPA、安全论证、验证确认 & 威胁建模、攻击树、渗透测试、访问控制、密码协议、安保论证 \\
典型证据 & 危害覆盖、失效率、诊断覆盖率、安全完整性、SOTIF 论证 & CIA 属性、威胁覆盖、攻击路径阻断、漏洞处置、安保控制有效性 \\
中文建议 & 安全、安全性；功能安全；预期功能安全 & 安保、安保性；网络安保；信息安保；数据安保 \\
\bottomrule
\label{table_summary_key_diff}
\end{tabularx}
\end{table}

\begin{longtable}{@{}p{0.18\textwidth}p{0.22\textwidth}p{0.34\textwidth}p{0.18\textwidth}@{}}
\caption{{safety} 与 {security} 在典型标准中的关注范围}\\
\toprule
概念/标准 & 主要对象 & 核心关注 & 本文建议译法 \\
\midrule
\endfirsthead
\toprule
概念/标准 & 主要对象 & 核心关注 & 本文建议译法 \\
\midrule
\endhead
ISO/IEC Guide 51 & 通用安全原则 & 免于不可容忍风险；风险由伤害概率与严重程度组合而成 & 安全性 \\
IEC 61508 / GB/T 20438 & E/E/PE安全相关系统 & 通过安全完整性、失效控制、验证确认实现功能安全 & 功能安全 \\
ISO 26262 / GB/T 34590 & 道路车辆 E/E 系统 & 面向汽车场景的功能安全生命周期、危害分析和风险控制 & 功能安全 \\
ISO 21448 / GB/T 43267 & 自动驾驶和先进驾驶辅助相关功能 & 无故障情况下由功能不足、性能局限或可预见误用导致的风险 & 预期功能安全 \\
ISO/IEC 27000 / 27001 & 信息资产与管理体系 & 保持保密性、完整性、可用性，并建立管理体系 & 信息安保 \\
NIST CSF 2.0 & 组织网络风险治理 & 识别、保护、检测、响应、恢复网络安保风险 & 网络安保 \\
ISO/SAE 21434 & 道路车辆网络安全工程 & 全生命周期识别和管理车辆网络安保风险 & 车辆网络安保 \\
UNECE R155/R156, ISO 24089 & 车辆网络安保与软件更新 & 网络安保管理体系、软件更新和更新管理体系 & 车辆安保/软件更新安保 \\
IEC 62443 & 工业自动化与控制系统 & 工控系统资产、区域、通道、角色和安保控制 & 工控系统安保 \\
GB/T 25069, GB/T 22239 等 & 国内信息安全、网络安全和等级保护体系 & 信息安全术语、等级保护、个人信息和车联网相关安全要求 & 信息安保/网络安保 \\
\bottomrule
\label{table_standards}
\end{longtable}

\subsection{简洁视角：“系统伤害环境”与“环境伤害系统”}

\begin{figure}[ht]
    \centering
    \IfFileExists{pic/ssloop.png}{%
      \includegraphics[width=0.8\linewidth]{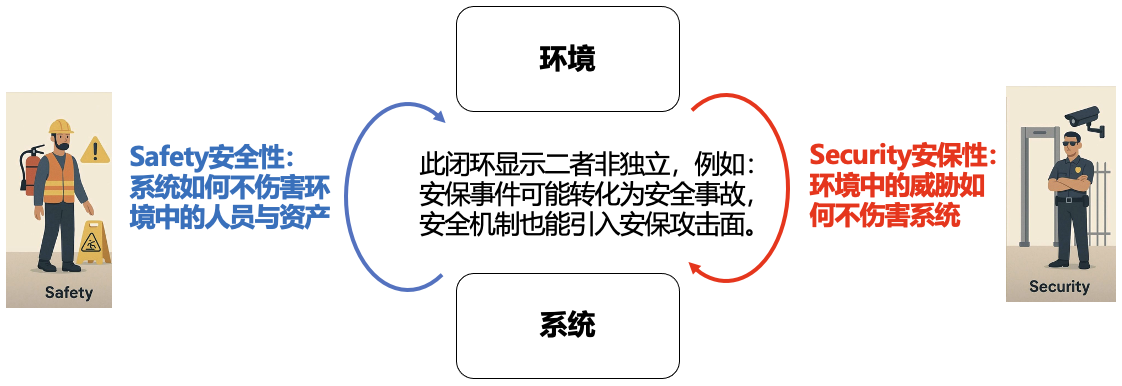}%
    }{%
      \missingfigure{Figure file \texttt{pic/ssloop.png} not found. The Overleaf/arXiv version will use the original safety-security loop image if the file is present.}%
    }
    \caption{区别安全与安保的一种简洁视角：“系统伤害环境”与“环境伤害系统”}
    \label{fig_ssloop}
\end{figure}

安全与安保二者的关键差异在于风险模型。如图\ref{fig_ssloop}所示，Bloomfield 与 Rushby在分析AI安全框架时给出的一个简洁视角：在关键系统的经典用法中，safety可被看作系统对环境可能造成的有害影响，security 则可被看作环境对系统可能造成的有害影响\parencite{bloomfieldRushbyWhereAIWrong2024}。这不是完备定义，却很好地揭示了二者的方向性：安全性关心“系统不要伤害环境”，安保性关心“环境中的威胁不要伤害系统”，此闭环显示二者非独立，安保事件可能转化为安全事故，安全机制也能引入安保攻击面。而复杂系统中的风险常常发生在这两个方向的交织处。例如，一次网络攻击可能导致列车制动失效、工业控制装置误动作或医疗设备剂量错误，从而把安保事件转化为安全事故。反过来，安全机制也可能引入新的安保攻击面，例如远程诊断接口、紧急维护口令和 OTA 更新链路。正因如此，学界已经发展出 safety-security co-assurance、security-informed safety、STPA-Sec、FMVEA 等联合分析方法\parencite{johnsonKelly2018Coassurance,bloomfieldNetkachovaStroud2013SIS,youngLeveson2014Integrated,schmittner2014FMVEA,schmittner2015FMVEAChassis}。联合分析的前提不是混同概念，而是先把概念分开，再研究它们如何耦合。

\subsection{AI assurance 将术语问题转化为论证问题}
\label{sec:ai-assurance}

上一节说明，{safety}与{security} 在标准体系中已经对应不同的风险模型、分析方法和证据类型；但在软件密集型系统、智能网联汽车和人工智能系统中，二者又经常相互影响。因此，问题并不是要把安全与安保割裂开，而是要在它们发生交互之前先让二者可区分。人工智能保障（AI assurance）\footnote{本文将{AI assurance} 暂译为“AI 保障”或“人工智能保障”，指通过测试、评估、审计、认证、论证和治理机制，对AI系统可信属性的关键主张（claim）进行度量、评价并向利益相关方沟通的活动。该译法并非现行中国国家标准中的固定译名；国内现行人工智能术语标准 GB/T 41867--2022《信息技术 人工智能 术语》尚未给出可作为通行规范的 {AI assurance} 对应译名。} 的前沿研究恰好把这一点推到台前：当一个含 AI 的系统需要被审查、认证、部署或监管时，争论的核心不再只是“这个系统安全吗”，而是具体哪一个主张、在什么系统边界内、针对哪类风险、由什么证据支撑\parencite{paterson2025SafetyAssuranceML,dong2023ReliabilityAssessmentML,bloomfieldRushbyAIDependability2025}。

这一点可以从 Bloomfield 与 Rushby 近年关于AI保障的工作中得到清晰说明。其关键启发并不是某个单一技术方案，而是一个系统工程视角：高风险系统不应把信任集中押在难以解释的 AI/ML 部件本身，而应通过防御纵深、多样性、运行时监控、外部守护器和更简单、更可论证的工程部件来降低对 AI/ML 部件的信任负担\parencite{bloomfieldRushbyAIDependability2025}。因此，AI assurance 的对象不应从“安全地使用 AI”（ {safe use of AI}）滑向抽象的“安全 AI”（ {safe AI}），而应回到部署系统、应用场景、运行域、系统边界和风险所有者\parencite{bloomfieldRushbyAssure2025}。这与 NIST AI RMF、ISO/IEC 23894 和 ISO/IEC 42001 等风险管理与治理框架形成互补关系：后者说明组织应管理什么，前者进一步追问这些要求如何在具体系统中变成可审查的主张（claims）、论证（arguments）、证据（evidence）和反驳项（defeaters）\parencite{nistAIRMF2023,isoIec23894,isoIec42001,bloomfieldRushbyModels2024,bloomfieldRushbyWhereAIWrong2024}。
AI 保障的最新方法论 Assurance 2.0 \parencite{bloomfieldRushbyAssurance2Manifesto2021} 进一步说明，术语必须支撑可质疑的论证。保障论证的核心不是写一份笼统说明，而是把顶层主张、推理结构、证据、理论假设和可能击败该主张的击败项组织起来，使开发者、评估者、监管者和公众能够审查其可信度\parencite{bloomfieldRushbyAssurance2Manifesto2021,bloomfieldRushbyConfidence2024,bloomfieldRushbyQuantifying2026,rushbyAssurance2Home}。如果 {safety} 主张与 {security} 主张都被写成“安全主张”，那么一个击败项到底是在质疑危害分析、失效假设、威胁模型、攻击能力、访问控制，还是供应链边界，就必须依赖上下文猜测而引起歧义和监管风险。举一个车辆例子：某自动驾驶安全论证声称“系统在所有识别出的危害场景下都有缓解措施”；一个潜在击败项可能是攻击者绕过 OTA 签名验证并植入改变感知阈值的软件更新。这首先是安保威胁，却会击败原本的安全主张。若中文评审材料只把二者都写成“安全问题”，评审者可能只追问 HARA 是否覆盖充分，而不会要求补充威胁模型、签名验证证据和供应链安保控制。此时，击败项机制并非不存在，而是被含混术语降低了被提出的概率。

\subsection{安全-安保协同论证不是把两个词合并}

安全与安保确实需要联合分析，但前沿研究恰恰表明，联合的前提是保留差异。Johnson 与 Kelly 提出的 Safety-Security Assurance Framework（SSAF）强调，复杂系统中的安全性与安保性协同保障困难，原因之一正是过程、信息、语言和专业哲学不匹配；他们主张让 安全 与 安保 保持相对独立的 保障 活动，再通过同步机制交换正确信息\parencite{johnsonKelly2018Coassurance}。换言之，协同保障不是把两个学科揉成一个“安全学科”，而是有组织地管理两个不同关注点之间的接口。

同样，“security-informed safety”\parencite{bloomfieldNetkachovaStroud2013SIS} 的口号“if it is not secure, it is not safe”并不意味着  {security} 与  {safety} 是同一个概念。Bloomfield等的原始工作基于结构化安全论证，讨论 安保性 对既有 安全性 的影响，以及如何评估安全相关系统中的 安保 风险\parencite{bloomfieldNetkachovaStroud2013SIS}。英国 NPSA 后续指南也强调，安全论证若不考虑安保影响就是不完整的\parencite{npsaSecurityInformedSafety}。这里的逻辑是“安保问题可能击败安全论证”，不是“安保问题就是安全问题”。若中文只说“安全影响安全论证”，论证关系本身就被抹平了。

近年的系统综述也显示，safety-security interrelations 已成为物理信息系统和关键基础设施研究中的快速增长主题；已有研究更多关注 安保性 对 安全性 的影响，但仍缺少显式处理二者相互关系的模型\parencite{zimmermann2025Interrelations}。这进一步说明，“安保性”不是为了割裂安全与安保，而是为了让“安全-安保关系”成为中文里可命名、可检索、可建模的研究对象。

\section{中文“安全”的过载}

\subsection{大类 safety 无法被干净表达}

在中文里，functional safety 已经稳定译作“功能安全”，SOTIF 已经译作“预期功能安全”。这两个译名在标准体系中具有现实约束，短期内不可能也不应轻易改动。但正因为如此，当我们想表达上位概念 {safety} 时，中文写作反而尴尬：若也叫“功能安全”，就与 {functional safety} 重合；若叫“系统安全”，又可能与 {system security} 或传统“系统安全工程”混淆；若只叫“安全”，又和 {security} 完全重合。

“安全性”可以作为 {safety} 的属性译法，但它仍然会与“信息安全性”“网络安全性”冲突。因此，更可行的策略是保留 {safety} 的“安全/安全性”核心地位，同时把 {security} 从“安全”里移出去。也就是说，不是给 {safety} 另找一个冷僻词，而是给 {security} 一个独立、更准确和通用的译名。

\subsection{Security 不是 information security 加 cybersecurity}

中文计算机领域常见一种隐含处理方式：将 {information security} 译作“信息安全”，将 {cybersecurity} 译作“网络安全”，但遇到上位词 {security} 时仍译作“安全”。这样做会带来一个逻辑问题：上位概念到底是什么？如果说 {security} 等于“信息安全加网络安全”，那么物理安保、供应链安保、身份与访问控制、产品安保、模型安保、操作技术安保和社会工程攻击应放在哪里？如果说 {security} 仍然就是“安全”，那么它又与 {safety} 无法区分。

“安保性”作为上位词的好处在于，它允许我们把 {information security}、 {cybersecurity}、 {data security}、 {product security}、 {model security}、 {systems security engineering} 放进同一个家族，而不必每次退回含混的“安全”。

% 例如：
% \begin{itemize}
%   \item {security}：安保性，或在领域名中简写为“安保”；
%   \item {information security}：信息安保；
%   \item {cybersecurity}：网络安保；
%   \item {data security}：数据安保；
%   \item {security engineering}：安保工程；
%   \item {security risk}：安保风险；
%   \item {security case}：安保论证；
%   \item {secure-by-design}：安保内建。
% \end{itemize}

\subsection{标准名称的既成事实反而说明问题}

国内标准已经把许多 {security} 译作“安全”。GB/T 25069 叫“信息安全技术 术语”，GB/T 22239 叫“信息安全技术 网络安全等级保护基本要求”，GB/T 40855、GB/T 40856、GB/T 40857 的中文名也使用“信息安全”，而英文名分别使用 {cybersecurity}\parencite{gbt25069,gbt22239,gbt40855,gbt40856,gbt40857}。法律层面，“网络安全法”“数据安全法”和“个人信息保护法”也已把相关 {security} 议题固定在“安全/保护”的中文制度语汇中\parencite{cybersecurityLawChina2016,dataSecurityLawChina2021,piplChina2021}。这不是本文要否认的事实，而恰恰是本文的问题意识：一个词已经承担了过多标准族的翻译任务。因此，本文倡议采取“双轨制”而不是“清零式改名”。引用既有标准和法律时，应忠实使用官方名称，例如“信息安全”“网络安全”“数据安全”等；在作者自己的概念论述中，可在首次出现时加注说明：“信息安全（{information security}；本文在概念分类中称为信息安保）”“网络安全（{cybersecurity}；本文在概念分类中称为网络安保）”。采用这一写法的后续研究可在术语说明处引用本文，从而不必在每篇文章中重复完整的翻译论证。这样既保留正式名称，又为后续分析提供稳定的术语区分。经过若干年共用，学术共同体才可能判断“安保性”是否足够自然、稳定和可扩展。

至此，本文只得到一个必要结论：{security} 不宜继续无条件地被上位“安全”吸收。但独立出来并不意味着可以任意造词。下一步问题是：既然需要一个中文词根承载 {security}，为什么选择“安保”而不是“安防”“防护”或其他说法？

\section{为什么是“安保性”}

{Security} 需要独立译名，是前文回答的问题；这个独立译名为什么应优先考虑“安保”，是本节要回答的问题。可选译法并不少。有人可能建议把 {security} 译为“安防”，因为“网络安防”“工控安防”听起来较顺；也有人可能建议用“保护性”“防护性”“保障性”。这些译法各有优点，但也有明显局限。

“安防”在中文里常与物理防盗、视频监控、门禁和公共安全产业绑定，放到密码学、访问控制、形式化验证和模型权重保护时，技术外延显得偏窄。“保密性”只对应 confidentiality，无法覆盖完整性、可用性、身份鉴别、不可否认性和供应链防护。“安全防护”可以解释概念，却不适合作为术语词根、不够凝练。

“安保”有三个优势。第一，它已经在公共治理和国际关系中承载 {security} 的保护含义，如前文提到的“安保机制”“安保条约”。第二，它可以自然派生：“安保性”“安保工程”“安保风险”“网络安保”“信息安保”。第三，它与“安全”相近但不重合，能在读者已有语感上建立新分工。

% 本文建议区分“安保”和“安保性”。当 {security} 表示一种系统属性或质量属性时，译作“安保性”；当它表示一个领域、实践或制度安排时，译作“安保”。例如：

% \begin{itemize}
%   \item The security of the protocol is proved. 译作“该协议的安保性得到证明”。
%   \item We study security engineering for autonomous vehicles. 译作“我们研究自动驾驶车辆的安保工程”。
%   \item Safety and security are both emergent system properties. 译作“安全性和安保性都是系统涌现属性”。
% \end{itemize}

% 这样处理也能避免“安保性工程”之类笨重表达。术语改革不应只追求字面对等，更要考虑中文句法中的可写性。

\section{建议术语表}

下表\ref{table_term}给出本文建议的最小术语集。它不是国家标准意义上的规范性定义，而是供论文、教材、项目文档和评审材料试用的写作约定。

\begin{longtable}{>{\raggedright\arraybackslash}p{0.24\textwidth}>{\raggedright\arraybackslash}p{0.22\textwidth}>{\raggedright\arraybackslash}p{0.42\textwidth}}
\caption{建议术语表}\\
\toprule
英文术语 & 建议中文 & 说明 \\
\midrule
\endfirsthead
\toprule
英文术语 & 建议中文 & 说明 \\
\midrule
\endhead
{safety} & 安全；安全性 & 指免于不可容忍风险的属性，尤其是对人员、财产、环境或任务造成伤害的风险。\\
{system safety} & 系统安全 & 可保留既有译法；必要时标注 {system safety} 以区别 {system security}。\\
{functional safety} & 功能安全 & 已固定于 IEC 61508、ISO 26262、GB/T 20438、GB/T 34590 等标准。\\
{SOTIF} & 预期功能安全 & 已固定于 ISO 21448 和 GB/T 43267。\\
{safety risk} & 安全风险 & 由危害、失效、功能不足或可预见误用导致的风险。\\
{safety case} & 安全论证 & 说明系统达到可接受/可容忍安全风险的结构化证据。\\
{security} & 安保；安保性 & 表示属性时用“安保性”，表示领域或实践时用“安保”。\\
{information security} & 信息安保 & 首次出现可写“信息安全（本文称信息安保）”。\\
{cybersecurity} & 网络安保 & 首次出现可写“网络安全（本文称网络安保）”。\\
{data security} & 数据安保 & 与法律或标准名并列时保留官方“数据安全”。\\
{AI safety} & 人工智能安全 & 可指 AI 模型/部件本身的安全属性，也可指在具体系统中安全地使用 AI；二者张力见第 \ref{sec:ai-assurance} 节。\\
{AI security} & 人工智能安保 & 主要指模型、数据、系统和部署链路抵御威胁、滥用和对抗攻击的能力。\\
{security risk} & 安保风险 & 来自威胁、脆弱性、攻击面、滥用或未授权行为的风险。\\
{security case} & 安保论证 & 说明系统满足安保目标和控制有效性的结构化证据。\\
{secure-by-design} & 安保内建 & 与 {safe-by-design}“安全内建”区分。\\
{safety-security co-analysis} & 安全-安保联合分析 & 研究安全性与安保性之间的相互影响。\\
{safety-security co-assurance} & 安全-安保协同保障 & 保持两个属性和专业过程的差异，同时管理其证据与风险接口。\\
{security-informed safety} & 安保知情安全；安保约束下的安全 & 指把安保威胁、攻击面和安保控制作为安全论证的重要输入。\\
{safety-informed security} & 安全知情安保；安全约束下的安保 & 指把功能安全机制、危害后果和安全完整性要求纳入安保分析。\\
{AI assurance} & AI 保障；人工智能保障（暂译） & {assurance} 尚无稳定中文译名；本文在需中文指称时暂用“AI 保障”或“人工智能保障”，并在必要处保留 {AI assurance}。具体论证结构可称为“保障论证”或“保证论证”；顶层主张为安全性或安保性时分别称“安全论证”或“安保论证”。\\
{safe use of AI} & 安全地使用 AI & 强调部署系统、运行域、用户、组织流程和应用情境中的安全性，是第 \ref{sec:ai-assurance} 节所主张的系统工程焦点。\\
{safe AI} & 安全 AI & 强调 AI 模型或部件本身的安全属性；若不说明系统语境，容易掩盖与 {safe use of AI} 的差别。\\
\bottomrule
\label{table_term}
\end{longtable}

\section{可能的反对意见与回应}

\subsection{反对一：约定俗成已经无法改变}

这是最强的反对意见。网络安全、信息安全、数据安全、等级保护、功能安全、预期功能安全都已经写入法律、标准、教材、学科方向和产业资质。本文并不主张把这些名称立即替换掉。相反，本文主张从增量空间做起：新论文、新教材、新术语表、新课程和新项目文档可以先采用“官方名 + 本文术语”的双写法，同时在术语说明处引用本文（来避免重复的翻译论证）。

更重要的是，既有术语已经进入标准、教材和工作流，并不只是改变困难的理由，也正是改变迫切的理由。按照 Bowker 与 Star 对分类基础设施的分析，一旦分类进入组织流程，它就会塑造记录、检索、审计和协作的能力\parencite{bowkerStar1999SortingThingsOut}。如果“安全”持续同时承载 {safety} 与 {security}，这种含混不会停留在口头语境，而会沉入需求模板、评审清单、数据库字段和学科目录。上下文解释只能补救单次沟通，不能修复基础设施层面的分类过载。

语言改革很少靠命令完成，而是靠高频、低摩擦、能解决真实问题的用法扩散。因此，倡议的目标不是要求所有既有名称立刻改写，而是尽快创造一组可试用、可引用、可检索的替代说法。若“安保性”确实能减少跨领域沟通成本，它会逐步获得生命；若不能，它会自然被淘汰。但在试用之前，中文学术共同体甚至缺少一个稳定候选项来承载这个区分。

\subsection{反对二：“安保”听起来不像计算机术语}

这个反对意见成立，但并不致命。许多今天自然的技术词最初都不像本领域术语。“防火墙”曾是建筑术语，“沙箱”来自儿童游戏和隔离环境，“云”来自隐喻图形。“安保”听起来偏公共治理，恰恰有助于提醒读者：{security} 不是狭义密码算法或漏洞扫描，而是保护资产、组织、任务和社会秩序的一整套工程活动。

此外，人工智能、智能网联汽车和工业控制系统已经把计算机系统带回公共空间。模型泄露、提示注入、供应链投毒、车端入侵、工控勒索和关键基础设施攻击，都不只是“信息技术内部问题”。“安保”一词的公共性反而是优点。

\subsection{反对三：safety 与 security 本来就交织，何必强分}

交织不是混同。医学中的病因、症状和治疗也交织，但仍需区分概念。安全性与安保性在 信息物理系统 中常相互影响，正因如此才需要清晰术语。“security-informed safety”、“safety-security co-assurance” 和 “integrated safety and security” 等新概念的出现，恰恰说明前沿研究不是放弃区分，而是在区分之后研究接口、击败关系和协同证据。没有清晰术语，联合分析就会退化。

\section{一套可执行的写作规范}

为了降低采用成本，本文建议中文科技写作采用以下七条规则。

\begin{enumerate}
  \item 首次出现 {safety} 或 {security} 时保留英文括注。例如“安全性（{safety}）”“安保性（{security}）”。
  \item 引用法律、标准和机构正式名称时不擅自改名。例如仍写“GB/T 22239《信息安全技术 网络安全等级保护基本要求》”，随后可说明“本文称其所属领域为信息安保/网络安保”。
  \item 在标题、摘要和关键词中避免单独使用“安全”指代 {security}。若论文其实讨论漏洞、攻击、认证、访问控制或威胁建模，优先写“安保性”。
  \item 同时讨论 {safety} 与 {security} 时，统一写作“安全性与安保性”，或在名词短语中写“安全-安保联合分析”。
  \item 对既有通用词保持宽容。公众传播可以继续使用“网络安全”；学术正文应在关键处给出精确术语。
  \item 后续作者若采用“安全性/安保性”区分，可在首次术语说明处引用本文，从而避免每篇文章都重复展开 {safety}/{security} 的翻译论证，并帮助这一用法积累为可检索、可复用的中文学术语料。
  \item 在双语术语表、数据集标签、课程大纲和审稿意见中积累用例。术语改革需要可检索的语料，而不只是立场宣言。
\end{enumerate}

\section{结论}

{Safety} 与 {security} 都重要，也都会伤人、毁坏财产、破坏信任、扰乱社会。但二者的工程逻辑不同：一个从危害和不可容忍风险出发，一个从资产、威胁和保护机制出发；一个主要问系统如何避免伤害环境，一个主要问环境中的威胁如何不伤害系统。中文把二者都译作“安全”，在许多场合曾经足够；但在软件密集、网络互联、物理闭环和人工智能驱动的新系统中，它正在变得不够。

本文倡议将 {security} 译作“安保”或“安保性”，不是为了追求陌生化，而是为了让中文科学与工程语言重新获得区分能力。更准确的术语会改变我们能提出什么问题、能建立什么证据、能审查什么论证。我们不必否认“网络安全”“信息安全”等既有名称，也不必期待标准体系一夜之间改写。可行的第一步很小：在自己的下一篇论文、下一份需求文档、下一张术语表里，把 {safety} 写成“安全性”，把 {security} 写成“安保性”。此安全非彼安全，先从说清楚开始。

\clearpage
\part*{English Version}
\addcontentsline{toc}{part}{English Version}

\begin{center}
{\LARGE\bfseries Not All \textit{anquan} Is the Same:\\A Terminological Proposal for Chinese Computer Science and Engineering\par}
\vspace{0.8em}
{\large Xingyu Zhao\par}
{\small School of Cyber Science and Engineering, Wuhan University\par}
{\small \texttt{xingyu.zhao@whu.edu.cn}\par}
\end{center}

\noindent\textbf{Note on the bilingual arXiv version.} The Chinese version appears first because the terminology proposal is primarily addressed to Chinese scientific and engineering writing. A complete English version follows for arXiv moderation, indexing and international readers.

\begin{abstract}
In Chinese computer science and engineering, \en{safety} and \en{security} have long been translated by the same word, \zh{安全} (\en{anquan}). This convention is concise in ordinary communication, but it creates persistent conceptual compression in standards interpretation, interdisciplinary collaboration, risk analysis and academic writing. When researchers need to discuss both whether a system is free from intolerable non-adversarial harm and whether it can resist adversarial threats, the single word \zh{安全} often cannot carry the distinction. This article argues that, while established legal and standards titles should be retained, scholarly and engineering writing should translate \en{security} as \zh{安保} or \zh{安保性} (\en{anbao/anbao-xing}), and reserve \zh{安全}/\zh{安全性} mainly for \en{safety}. This is not a cosmetic translation preference, but a proposal for terminological governance in scientific cognition, engineering risk communication and assurance argumentation. The article first surveys the conceptual boundary between \en{safety} and \en{security} in international and Chinese standards, and analyzes how the current translation overload affects functional safety, SOTIF, information security, cybersecurity, automotive cybersecurity and AI governance. It then uses recent work on AI assurance, safety-security co-assurance and security-informed safety to show why precise terminology is fundamental to scientific arguments that can be examined, challenged and communicated. Finally, it proposes a staged, dual-track writing practice for Chinese technical discourse.
\end{abstract}

\noindent\textbf{Keywords:} terminology translation; safety; security; \zh{安全}; \zh{安保性}; AI assurance; Assurance 2.0; security-informed safety; functional safety; SOTIF

\section{Introduction}

On 2 November 2023, during the first AI Safety Summit, the UK government announced the creation of the AI Safety Institute, stating that its mission was to evaluate risks from frontier AI models \parencite{govuk2023AISIlaunch,govuk2023AISIoverview}. A little over a year later, on 14 February 2025, the government announced that the institution would be renamed the AI Security Institute, emphasizing national security, criminal misuse and related protective work as central to its mission \parencite{govuk2025AISIRename}. The acronym AISI remained unchanged, but the key term on the sign changed from \en{safety} to \en{security}. This example illustrates a simple point: in English-language science policy and engineering governance, \en{safety} and \en{security} are not interchangeable near-synonyms. Both are concerned with bad outcomes, but the sources of those outcomes, the risk models, the engineering evidence, the organizational responsibilities and the governance mechanisms differ. In Chinese technical discourse, however, computer science and engineering scholars have long translated both terms as \zh{安全}. As a result, AI safety, AI security, functional safety, SOTIF, cybersecurity, information security, system safety engineering and systems security engineering are all compressed into the same Chinese lexical root: \zh{安全}.

\begin{figure}[ht]
    \centering
    \IfFileExists{pic/aisi.png}{%
      \includegraphics[width=0.8\linewidth]{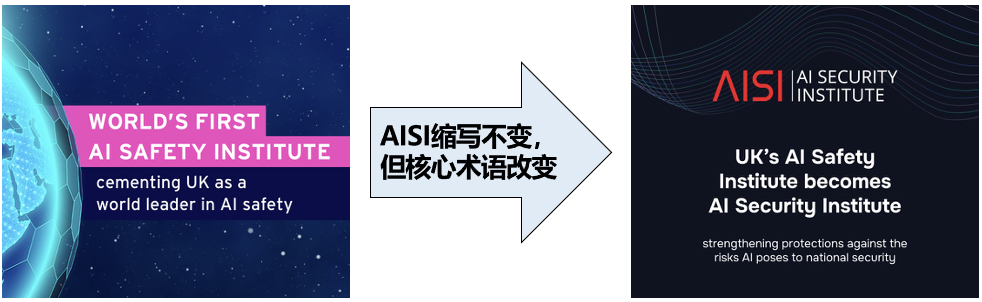}%
    }{%
      \missingfigure{Figure file \texttt{pic/aisi.png} not found. The Overleaf version will use the original AISI image if the file is present.}%
    }
    \caption{The AISI renaming exposes a conceptual boundary: the same government institution changed from \en{safety} to \en{security}, indicating that the two terms perform different conceptual roles in risk classification, governance boundaries and organizational missions.}
    \label{fig:aisi}
\end{figure}

The cost of this compression may sometimes be repaired by context within a single mature field. But in software-defined vehicles, industrial control, medical devices, robotics, aerospace and general-purpose AI, the interaction between \en{safety} and \en{security} has become a central problem. A vehicle perception system may have no random hardware fault, yet still create a safety risk because of performance limitations. The same system may create a security risk through remote compromise, data poisoning or authorization bypass. If both are simply called \zh{安全风险}, risk review meetings, requirements documents, paper titles and standards indexes become ambiguous. This semantic overlap and vagueness does not merely increase communication cost; it may also create misunderstanding in key concepts and mislead risk judgment and decision making.

This article therefore makes a deliberately somewhat uncomfortable proposal: in computer science and engineering research, \en{security} should be translated as \zh{安保} or \zh{安保性}, while \en{safety} should continue to be translated as \zh{安全} or \zh{安全性}. The word \zh{安保性} borrows from the use of \zh{安保} in international relations and public governance. For example, the Embassy of China in Peru, the Peruvian National Police and the Association of Chinese Enterprises in Peru established a ``China-Peru bilateral security liaison mechanism'' called the \zh{中秘双边安保联席会议机制} \parencite{mfa2022ChinaPeruAnbaoMechanism}. In that usage, \zh{安保} clearly refers to protective security arrangements for personnel, institutions and projects. Compared with \zh{安全}, \zh{安保} more directly evokes protection, safeguarding, guarding and adversarial threat response. This article does not propose immediately rewriting laws, national standards or entrenched industry names. It proposes that future papers, textbooks, glossaries and engineering documents in computer science and engineering give \en{security} a stable Chinese home through \zh{安保性}.

\section{Terminology and Cognition}

Terminological questions are easily dismissed as questions of wording. In science, however, terms are not labels attached to already-given objects. They are tools through which research communities identify objects, draw boundaries, organize evidence and allocate responsibility. Wittgenstein wrote in the \en{Tractatus} that the limits of language mean the limits of the world \parencite{wittgenstein1922Tractatus}. A moderate version of linguistic relativity need not claim that language determines thought; it is enough to note that lexical and categorical resources can become default paths for conceptualization and action \parencite{whorf1956Language,lucy1998SapirWhorf}. People can of course encounter an experience without having a term for it. But when a community lacks a stable term, it is harder to turn that experience into something that can be indexed, taught, reviewed and standardized.

This is the deeper problem with translating \en{safety} and \en{security} by the same word \zh{安全}. Chinese researchers are not unaware that the two concepts differ. The practical problem is that the public technical vocabulary does not give this difference a low-cost and durable place to live. As a result, the distinction must be reconstructed from context each time, rather than being stabilized in titles, keywords, course modules, standards indexes, requirement fields and review checklists. Bowker and Star's work on classification systems reminds us that once classifications enter workflows and infrastructure, they shape what organizations can see, ignore and record \parencite{bowkerStar1999SortingThingsOut}. In engineering, a terminological error is not merely a rhetorical blemish; it can become an error in risk registers, responsibility assignment and evidence structure.

Thus the proposed term \zh{安保性} is not intended to manufacture novelty. It is intended to provide a Chinese cognitive anchor for a concept family that already exists and is becoming increasingly important. Without that anchor, existing and emerging research areas such as \en{safety case} and \en{security case}, \en{AI safety} and \en{AI security}, \en{safety-security co-assurance} and \en{security-informed safety} repeatedly collapse into \zh{安全论证}, \zh{人工智能安全}, \zh{安全-安全协同保证} or \zh{安全知悉的安全}. Such collapse constrains how researchers formulate questions. If a language repeatedly folds two objects into one, researchers must first spend additional effort separating them again.

\section{Conceptual Boundaries}

Because classification systems shape what engineers can see, the boundary between \en{safety} and \en{security} is not an after-the-fact naming matter. It directly affects hazard analysis, threat modeling and the scope of assurance arguments. If the two are absorbed by the single Chinese term \zh{安全}, experienced engineers may still distinguish them locally, but documents, standards, tools and review procedures cannot record the distinction consistently. The following survey of standards shows that the conceptual difference already corresponds to different problem structures, evidence types and divisions of responsibility.

In the classical taxonomy of dependability, \en{safety} and \en{security} are not two names for the same property. They are neighboring attributes in a broader network that includes reliability, availability, integrity and maintainability \parencite{avizienis2004Dependable}. They are adjacent, so they must be coordinated. They are distinct, so they should not be absorbed indiscriminately by the same Chinese word.

\subsection{Safety: Systems Should Not Harm People, the Environment or the Mission}

Intuitively, \en{safety} asks whether \textbf{a system, under its intended use and reasonably foreseeable misuse, avoids causing intolerable harm to people, property, the environment or the mission}. Its typical question is not whether there is an attacker. It is how failures, functional insufficiencies, design defects, environmental disturbances, human-machine interaction errors or organizational process failures may bring a system into a hazardous state. In short, \en{safety} first asks: \textbf{will the system harm its environment?}

The typical risk sources for \en{safety} are therefore hazards, failure modes, functional insufficiencies, operational errors, foreseeable misuse and complex system interactions, rather than an intentional, goal-directed and adaptive adversary. Its engineering methods accordingly focus on hazard analysis, risk reduction, safety integrity, verification and validation, and safety cases. ISO/IEC Guide 51 defines \en{safety} as freedom from risk that is not tolerable, and understands risk as a combination of the probability of occurrence of harm and the severity of that harm \parencite{isoGuide51,kishimoto2015Guide51}. System safety research has developed systematic analyses of accident causation, organizational processes and control structures \parencite{leveson2012SaferWorld}.

In engineering, this tradition has produced a large standards family. IEC 61508 addresses the functional safety of electrical, electronic and programmable electronic safety-related systems, with software safety requirements forming an important part \parencite{iec61508-1,iec61508-3}. The Chinese GB/T 20438 series corresponds to this functional safety family \parencite{gbt20438-1}. ISO 26262 extends functional safety to road vehicle E/E systems, and GB/T 34590 adopts the Chinese term \zh{功能安全} \parencite{iso26262-1,gbt34590-1}. ISO 21448 and GB/T 43267 address Safety of the Intended Functionality, or SOTIF, where unreasonable risk may arise from functional insufficiencies, performance limitations or foreseeable misuse even without system failure \parencite{iso21448,gbt43267}. The common theme of these standards is not secrecy or intrusion prevention. It is achieving tolerable risk through hazard analysis, risk reduction, safety integrity, verification and validation, and safety arguments.

\subsection{Security: Systems and Assets Should Not Be Damaged by Threats}

By contrast, \en{security} asks whether \textbf{systems, data, assets, services and missions are protected against external or internal threats, attacks, misuse, unauthorized access and adversarial behavior}. If \en{safety} first asks whether the system harms the environment, then \en{security} first asks: \textbf{will threats in the environment harm the system, its assets or its mission?} In the remainder of this article, \en{security} is translated as \zh{安保} or \zh{安保性}.

The core concern of security is not random failure or performance insufficiency as such. It is assets, threat actors, attack surfaces, vulnerabilities, privileges, authentication, access control, supply chains, data integrity and service availability. Strictly speaking, ISO/IEC 27000 defines \en{information security}, namely the preservation of confidentiality, integrity and availability of information, while noting that authenticity, accountability, non-repudiation and reliability may also be involved. ISO/IEC 27001 institutionalizes these objectives as requirements for an information security management system \parencite{isoIec27000,isoIec27001}. Saltzer and Schroeder's classic paper on information protection discusses how to prevent unauthorized use or modification \parencite{saltzer1975Protection}. Classical security engineering textbooks are organized around threats, access control, protocols, software vulnerabilities and system protection \parencite{anderson2020SecurityEngineering,bishop2018ComputerSecurity}. NIST Cybersecurity Framework 2.0 addresses organizational management of cybersecurity risk \parencite{nistCSF2024}, while NIST SP 800-160 treats trustworthy secure systems and cyber-resilient systems from a systems engineering perspective \parencite{nistSP800160v1r1,nistSP800160v2r1}.

Security analysis has also developed method families different from hazard analysis. Misuse cases and attack trees make requirement gaps and attack paths explicit \parencite{sindre2005MisuseCases,schneier1999AttackTrees}. In sectoral standards, ISO/SAE 21434 addresses road vehicle cybersecurity engineering and requires cybersecurity risk management across the lifecycle from concept, development and production to operation, maintenance and decommissioning \parencite{isoSae21434}. UNECE R155/R156 and ISO 24089 further institutionalize vehicle cybersecurity management, software updates and software update management systems \parencite{unR155,unR156,iso24089}. The IEC 62443 series addresses security for industrial automation and control systems, covering asset owners, system integrators and product suppliers \parencite{iec62443-1-1,iec62443-2-1}. In China, GB/T 25069 \en{Information security technology: Terminology}, GB/T 22239 on classified protection of cybersecurity, GB/T 35273 on personal information security specification, and standards on automotive gateways, vehicle information interaction and electric vehicle remote service systems all fall into this family \parencite{gbt25069,gbt22239,gbt35273,gbt40855,gbt40856,gbt40857}. Table~\ref{table_summary_key_diff_en} summarizes the key engineering differences, and Table~\ref{table_standards_en} summarizes typical safety and security standards and their scope.

\begin{table}[htbp]
\centering
\caption{Typical engineering differences between \en{safety} and \en{security}}
\label{table_summary_key_diff_en}
\begin{tabularx}{\textwidth}{>{\raggedright\arraybackslash}p{0.18\textwidth}XX}
\toprule
Dimension & Safety: \zh{安全性} & Security: \zh{安保性} \\
\midrule
Main question & How the system avoids causing intolerable harm & How the system protects assets against threats, misuse and unauthorized behavior \\
Risk sources & Hazards, failures, performance limits, environmental disturbances, foreseeable misuse, organizational deficiencies & Attackers, misuse, insiders, malware, supply chains, unauthorized access, adversarial examples \\
Objects of analysis & Hazards, hazardous states, accident scenarios, risk reduction measures & Assets, threats, vulnerabilities, attack surfaces, controls \\
Typical methods & HARA, FMEA, FTA, STPA, safety cases, verification and validation & Threat modeling, attack trees, penetration testing, access control, cryptographic protocols, security cases \\
Typical evidence & Hazard coverage, failure rates, diagnostic coverage, safety integrity, SOTIF arguments & CIA properties, threat coverage, blocked attack paths, vulnerability handling, effectiveness of security controls \\
Chinese terms proposed & \zh{安全}, \zh{安全性}; \zh{功能安全}; \zh{预期功能安全} & \zh{安保}, \zh{安保性}; \zh{网络安保}; \zh{信息安保}; \zh{数据安保} \\
\bottomrule
\end{tabularx}
\end{table}

\begin{longtable}{@{}p{0.18\textwidth}p{0.22\textwidth}p{0.34\textwidth}p{0.18\textwidth}@{}}
\caption{The scope of \en{safety} and \en{security} in typical standards \label{table_standards_en}}\\
\toprule
Concept/standard & Main object & Core concern & Proposed Chinese term \\
\midrule
\endfirsthead
\toprule
Concept/standard & Main object & Core concern & Proposed Chinese term \\
\midrule
\endhead
ISO/IEC Guide 51 & General safety principles & Freedom from intolerable risk; risk combines probability and severity of harm & \zh{安全性} \\
IEC 61508 / GB/T 20438 & E/E/PE safety-related systems & Functional safety through safety integrity, failure control, verification and validation & \zh{功能安全} \\
ISO 26262 / GB/T 34590 & Road vehicle E/E systems & Functional safety lifecycle, hazard analysis and risk control for automotive systems & \zh{功能安全} \\
ISO 21448 / GB/T 43267 & Automated-driving and driver-assistance functions & Risk from functional insufficiencies, performance limits or foreseeable misuse without system failure & \zh{预期功能安全} \\
ISO/IEC 27000 / 27001 & Information assets and management systems & Confidentiality, integrity, availability and management-system requirements & \zh{信息安保} \\
NIST CSF 2.0 & Organizational cyber risk governance & Identify, protect, detect, respond and recover in cybersecurity risk management & \zh{网络安保} \\
ISO/SAE 21434 & Road vehicle cybersecurity engineering & Lifecycle identification and management of vehicle cybersecurity risks & \zh{车辆网络安保} \\
UNECE R155/R156, ISO 24089 & Vehicle cybersecurity and software update & Cybersecurity management systems, software updates and update management systems & \zh{车辆安保}/\zh{软件更新安保} \\
IEC 62443 & Industrial automation and control systems & Assets, zones, conduits, roles and security controls for industrial control systems & \zh{工控系统安保} \\
GB/T 25069, GB/T 22239 and related standards & Chinese information-security, cybersecurity and classified-protection systems & Information-security terminology, classified protection, personal information and vehicle-network requirements & \zh{信息安保}/\zh{网络安保} \\
\bottomrule
\end{longtable}

\subsection{A Concise View: ``Systems Harm the Environment'' and ``the Environment Harms Systems''}

\begin{figure}[ht]
    \centering
    \IfFileExists{pic/ssloop.png}{%
      \includegraphics[width=0.8\linewidth]{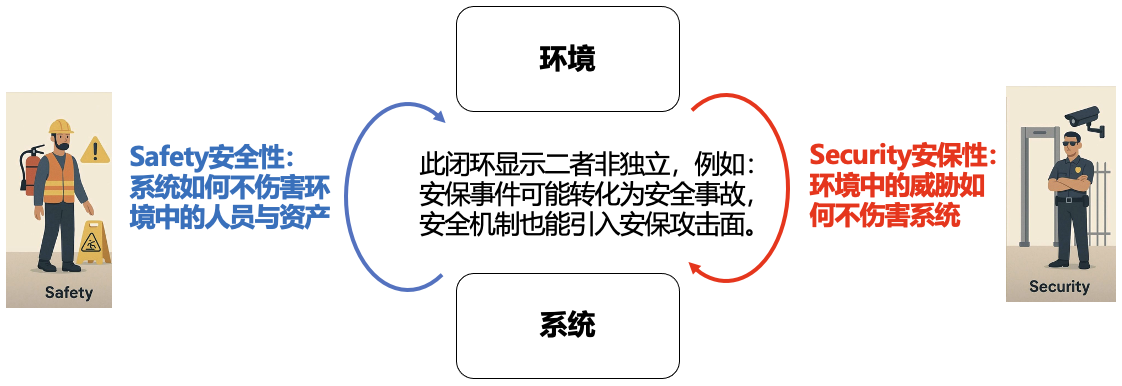}%
    }{%
      \missingfigure{Figure file \texttt{pic/ssloop.png} not found. The Overleaf version will use the original safety-security loop image if the file is present.}%
    }
    \caption{A concise view for distinguishing safety and security: ``systems harm the environment'' versus ``the environment harms systems.''}
    \label{fig:ssloop}
\end{figure}

The key difference between safety and security lies in their risk models. As shown in Figure~\ref{fig:ssloop}, Bloomfield and Rushby offer a concise view in their analysis of AI safety frameworks: in classical critical-system usage, \en{safety} can be understood as the harmful effect of the system on its environment, whereas \en{security} can be understood as the harmful effect of the environment on the system \parencite{bloomfieldRushbyWhereAIWrong2024}. This is not a complete definition, but it captures the directional difference. Safety asks that systems not harm their environment; security asks that threats in the environment not harm the system. The loop also shows that the two are not independent. A security event can become a safety accident, and a safety mechanism can introduce a security attack surface. For example, a cyberattack can cause train braking failure, industrial-control misoperation or incorrect dosage in a medical device. Conversely, safety mechanisms such as remote diagnostic interfaces, emergency maintenance passwords and OTA update channels can create new security attack surfaces. For this reason, researchers have developed joint analysis methods such as safety-security co-assurance, security-informed safety, STPA-Sec and FMVEA \parencite{johnsonKelly2018Coassurance,bloomfieldNetkachovaStroud2013SIS,youngLeveson2014Integrated,schmittner2014FMVEA,schmittner2015FMVEAChassis}. Joint analysis does not require conceptual merger. It requires separating the concepts first and then studying how they couple.

\subsection{AI Assurance Turns Terminology into an Argumentation Problem}
\label{sec:ai-assurance-en}

The previous section shows that \en{safety} and \en{security} already correspond to different risk models, analysis methods and evidence types in standards. Yet in software-intensive systems, connected vehicles and AI systems, the two also interact frequently. The point, therefore, is not to isolate safety and security, but to make them distinguishable before their interactions are analyzed. Recent work on AI assurance brings this issue to the foreground. When an AI-containing system must be reviewed, certified, deployed or regulated, the core question is no longer merely whether ``the system is safe.'' It is which claim is being made, within which system boundary, against which class of risk, and on what evidence \parencite{paterson2025SafetyAssuranceML,dong2023ReliabilityAssessmentML,bloomfieldRushbyAIDependability2025}.

Here this article provisionally renders \en{AI assurance} as \zh{AI 保障} or \zh{人工智能保障}.\footnote{This provisional Chinese translation refers to activities that measure, evaluate, audit, certify, argue for and communicate key claims about trustworthy properties of AI systems through testing, assessment, audit, certification, argumentation and governance mechanisms. It is not presented as a fixed term in current Chinese national standards. The Chinese AI terminology standard GB/T 41867--2022 \en{Information technology: Artificial intelligence terminology} does not, to the author's knowledge, provide a widely accepted normative translation of \en{AI assurance}.} The key lesson from Bloomfield and Rushby's recent work is not a single technical recipe, but a systems engineering perspective: high-risk systems should not concentrate trust in opaque AI/ML components; rather, they should reduce the trust burden placed on those components through defense in depth, diversity, runtime monitoring, external guards and simpler engineered components that are easier to justify \parencite{bloomfieldRushbyAIDependability2025}. Thus AI assurance should not slide from the \en{safe use of AI} toward the abstract idea of \en{safe AI}; it should return to deployed systems, application contexts, operating domains, system boundaries and risk owners \parencite{bloomfieldRushbyAssure2025}. This complements risk-management and governance frameworks such as the NIST AI RMF, ISO/IEC 23894 and ISO/IEC 42001. Those frameworks specify what organizations should manage; the assurance literature further asks how those requirements become reviewable claims, arguments, evidence and defeaters in concrete systems \parencite{nistAIRMF2023,isoIec23894,isoIec42001,bloomfieldRushbyModels2024,bloomfieldRushbyWhereAIWrong2024}.

Assurance 2.0 further shows that terminology must support challengeable arguments \parencite{bloomfieldRushbyAssurance2Manifesto2021}. The point of an assurance case is not to write a generic explanatory document, but to organize top-level claims, inference structures, evidence, theory assumptions and defeaters so that developers, assessors, regulators and the public can examine their credibility \parencite{bloomfieldRushbyAssurance2Manifesto2021,bloomfieldRushbyConfidence2024,bloomfieldRushbyQuantifying2026,rushbyAssurance2Home}. If a \en{safety claim} and a \en{security claim} are both rendered as \zh{安全主张}, then a defeater may be challenging hazard analysis, failure assumptions, threat models, attacker capability, access control or supply-chain boundaries, and the reader must infer which from context. That creates ambiguity and regulatory risk. Consider an automated-vehicle example: a safety case claims that the system has mitigations for all identified hazardous scenarios. A potential defeater is that an attacker can bypass OTA signature verification and install a software update that changes a perception threshold. This is first a security threat, but it defeats the original safety claim. If Chinese review materials call both matters \zh{安全问题}, reviewers may ask only whether HARA coverage is adequate, without requiring a threat model, signature-verification evidence or supply-chain security controls. In that case, the defeater mechanism has not disappeared; it has become less likely to be raised because the terminology hides it.

\subsection{Safety-Security Co-Assurance Does Not Mean Merging the Words}

Safety and security do need joint analysis, but the research frontier shows that joint analysis presupposes their difference. Johnson and Kelly's Safety-Security Assurance Framework (SSAF) emphasizes that integrated safety and security assurance in complex systems is difficult partly because the two communities differ in processes, information, language and professional philosophy. They propose keeping safety and security assurance activities relatively independent while using synchronization mechanisms to exchange the right information \parencite{johnsonKelly2018Coassurance}. In other words, co-assurance does not merge two disciplines into one generic \zh{安全} discipline; it manages the interfaces between two distinct concerns.

Likewise, the slogan of \en{security-informed safety}, ``if it is not secure, it is not safe,'' does not mean that \en{security} and \en{safety} are the same concept \parencite{bloomfieldNetkachovaStroud2013SIS}. Bloomfield and colleagues' original work uses structured safety cases to examine how security affects existing safety cases and how security risks in safety-related systems can be assessed \parencite{bloomfieldNetkachovaStroud2013SIS}. Later guidance from the UK's National Protective Security Authority also emphasizes that a safety case is incomplete if it does not consider security impacts \parencite{npsaSecurityInformedSafety}. The logic is that a security issue may defeat a safety argument, not that a security issue is simply a safety issue. If Chinese can only say \zh{安全影响安全论证}, the argumentative relation is flattened.

Recent systematic-review work also shows that safety-security interrelations have become a rapidly growing topic in cyber-physical systems and critical-infrastructure research. Existing work focuses especially on the influence of security on safety, but still lacks explicit models of their mutual relations \parencite{zimmermann2025Interrelations}. This further illustrates that \zh{安保性} is not intended to split safety from security. It is intended to make ``safety-security relations'' nameable, searchable and modelable in Chinese.

\section{The Overload of Chinese \zh{安全}}

\subsection{The Superordinate Category of Safety Cannot Be Expressed Cleanly}

In Chinese, \en{functional safety} is already fixed as \zh{功能安全}, and SOTIF is already fixed as \zh{预期功能安全}. These translations are institutionally constrained by standards and should not be lightly changed. Precisely because of this, however, Chinese writing becomes awkward when it needs to express the superordinate concept \en{safety}. If it is also called \zh{功能安全}, it collapses into \en{functional safety}. If it is called \zh{系统安全}, it may be confused with \en{system security} or the older Chinese usage of \zh{系统安全工程}. If it is simply called \zh{安全}, it collapses into \en{security}.

\zh{安全性} can serve as the property translation of \en{safety}, but it still conflicts with phrases such as \zh{信息安全性} and \zh{网络安全性}. A more workable strategy is therefore to preserve \zh{安全}/\zh{安全性} as the core rendering of \en{safety}, while moving \en{security} out of \zh{安全}. In other words, the task is not to find an obscure new term for \en{safety}, but to give \en{security} an independent, more accurate and more general Chinese term.

\subsection{Security Is Not Information Security Plus Cybersecurity}

A common implicit practice in Chinese computing is to translate \en{information security} as \zh{信息安全} and \en{cybersecurity} as \zh{网络安全}, but still translate the superordinate \en{security} as \zh{安全}. This creates a logical problem: what exactly is the superordinate category? If \en{security} means only information security plus cybersecurity, where do physical security, supply-chain security, identity and access control, product security, model security, operational-technology security and social-engineering attacks belong? If \en{security} still simply means \zh{安全}, then it remains indistinguishable from \en{safety}.

The benefit of \zh{安保性} as a superordinate term is that it can house \en{information security}, \en{cybersecurity}, \en{data security}, \en{product security}, \en{model security} and \en{systems security engineering} in the same family, without forcing writers to retreat each time to the ambiguous word \zh{安全}.

\subsection{Entrenched Standards Names Are Part of the Evidence}

Chinese standards already translate many \en{security} terms as \zh{安全}. GB/T 25069 is titled \zh{信息安全技术 术语}; GB/T 22239 is titled \zh{信息安全技术 网络安全等级保护基本要求}; and GB/T 40855, GB/T 40856 and GB/T 40857 also use \zh{信息安全} in Chinese while their English titles use \en{cybersecurity} \parencite{gbt25069,gbt22239,gbt40855,gbt40856,gbt40857}. In law, the Cybersecurity Law, Data Security Law and Personal Information Protection Law have also fixed related security issues within the Chinese institutional vocabulary of \zh{安全} and \zh{保护} \parencite{cybersecurityLawChina2016,dataSecurityLawChina2021,piplChina2021}. This article does not deny these facts. They are precisely the source of the problem: one word has been asked to carry too many standards families.

For this reason, the proposal is a dual-track practice, not a clean-slate renaming. When citing existing standards and laws, authors should faithfully use official titles such as \zh{信息安全}, \zh{网络安全} and \zh{数据安全}. In their own conceptual analysis, however, they can add a note at first use, for example: \zh{信息安全} (\en{information security}; in this article's conceptual classification, \zh{信息安保}) or \zh{网络安全} (\en{cybersecurity}; in this article's conceptual classification, \zh{网络安保}). Later studies that adopt this writing practice can cite this article in their first terminological note, so that each paper does not need to repeat the full translation argument. This approach preserves official names while giving later analysis a stable distinction. Only after years of use will the scholarly community be able to judge whether \zh{安保性} is natural, stable and extensible enough.

At this point the article has established a necessary conclusion: \en{security} should not continue to be unconditionally absorbed by the superordinate \zh{安全}. But separation does not justify arbitrary word creation. The next question is why the Chinese root should be \zh{安保} rather than \zh{安防}, \zh{防护} or some other alternative.

\section{Why \zh{安保性}?}

The previous sections answer why \en{security} needs an independent Chinese rendering. This section explains why \zh{安保} should be considered first. There are several possible alternatives. Some may prefer \zh{安防}, because expressions such as \zh{网络安防} or \zh{工控安防} sound familiar. Others may prefer \zh{保护性}, \zh{防护性} or \zh{保障性}. Each has merits, but each also has limitations.

\zh{安防} is strongly associated in Chinese with physical anti-theft systems, video surveillance, access-control equipment and public-safety industries. In cryptography, access control, formal verification and model-weight protection, its extension is too narrow. \zh{保密性} corresponds only to confidentiality and cannot cover integrity, availability, authentication, non-repudiation or supply-chain protection. \zh{安全防护} can explain the idea, but it is too heavy to function as a productive terminological root.

\zh{安保} has three advantages. First, it already carries the protective meaning of \en{security} in public governance and international relations, as in \zh{安保机制} and \zh{安保条约}. Second, it derives naturally: \zh{安保性}, \zh{安保工程}, \zh{安保风险}, \zh{网络安保}, \zh{信息安保}. Third, it is close enough to \zh{安全} to be intelligible but different enough to create a new division of labor.

\section{A Suggested Glossary}

Table~\ref{table_term_en} gives the minimum terminology set proposed by this article. It is not a national-standard definition. It is a writing convention for trial use in papers, textbooks, project documents and review materials.

\begin{longtable}{>{\raggedright\arraybackslash}p{0.24\textwidth}>{\raggedright\arraybackslash}p{0.22\textwidth}>{\raggedright\arraybackslash}p{0.42\textwidth}}
\caption{Suggested glossary \label{table_term_en}}\\
\toprule
English term & Suggested Chinese & Explanation \\
\midrule
\endfirsthead
\toprule
English term & Suggested Chinese & Explanation \\
\midrule
\endhead
\en{safety} & \zh{安全}; \zh{安全性} & Freedom from intolerable risk, especially risk of harm to people, property, the environment or missions.\\
\en{system safety} & \zh{系统安全} & Existing usage can be retained; mark \en{system safety} when confusion with \en{system security} is possible.\\
\en{functional safety} & \zh{功能安全} & Fixed in IEC 61508, ISO 26262, GB/T 20438 and GB/T 34590.\\
\en{SOTIF} & \zh{预期功能安全} & Fixed in ISO 21448 and GB/T 43267.\\
\en{safety risk} & \zh{安全风险} & Risk caused by hazards, failures, functional insufficiency or foreseeable misuse.\\
\en{safety case} & \zh{安全论证} & Structured evidence and argument that the system achieves acceptable or tolerable safety risk.\\
\en{security} & \zh{安保}; \zh{安保性} & Use \zh{安保性} for a property, and \zh{安保} for a field or practice.\\
\en{information security} & \zh{信息安保} & At first use, one may write \zh{信息安全（本文称信息安保）}.\\
\en{cybersecurity} & \zh{网络安保} & At first use, one may write \zh{网络安全（本文称网络安保）}.\\
\en{data security} & \zh{数据安保} & Retain the official \zh{数据安全} when citing laws or standards.\\
\en{AI safety} & \zh{人工智能安全} & May refer to safety properties of AI models/components or to the safe use of AI in concrete systems; see Section~\ref{sec:ai-assurance-en}.\\
\en{AI security} & \zh{人工智能安保} & Resistance to threats, misuse and adversarial attacks against models, data, systems and deployment chains.\\
\en{security risk} & \zh{安保风险} & Risk from threats, vulnerabilities, attack surfaces, misuse or unauthorized behavior.\\
\en{security case} & \zh{安保论证} & Structured evidence and argument that security objectives and controls are satisfied.\\
\en{secure-by-design} & \zh{安保内建} & Distinguishes it from \en{safe-by-design}, \zh{安全内建}.\\
\en{safety-security co-analysis} & \zh{安全-安保联合分析} & Analysis of mutual influence between safety and security.\\
\en{safety-security co-assurance} & \zh{安全-安保协同保障} & Coordinated management of evidence and risk interfaces while preserving the difference between the two properties and professional processes.\\
\en{security-informed safety} & \zh{安保知情安全}; \zh{安保约束下的安全} & Incorporating security threats, attack surfaces and security controls as important inputs to safety arguments.\\
\en{safety-informed security} & \zh{安全知情安保}; \zh{安全约束下的安保} & Incorporating functional-safety mechanisms, hazard consequences and safety-integrity requirements into security analysis.\\
\en{AI assurance} & \zh{AI 保障}; \zh{人工智能保障} (provisional) & \en{Assurance} has no stable Chinese translation. This article uses \zh{AI 保障} provisionally and keeps \en{AI assurance} where helpful.\\
\en{safe use of AI} & \zh{安全地使用 AI} & Emphasizes safety in deployed systems, operating domains, users, organizational processes and application contexts.\\
\en{safe AI} & \zh{安全 AI} & Emphasizes safety properties of AI models or components; without system context, it can obscure the distinction from \en{safe use of AI}.\\
\bottomrule
\end{longtable}

\section{Objections and Replies}

\subsection{Objection 1: Established Usage Cannot Be Changed}

This is the strongest objection. \zh{网络安全}, \zh{信息安全}, \zh{数据安全}, classified protection, \zh{功能安全} and \zh{预期功能安全} have entered laws, standards, textbooks, discipline names and industrial qualifications. This article does not propose replacing these names immediately. It proposes starting in incremental spaces: new papers, textbooks, glossaries, courses and project documents can use a dual form, combining official names with the terms proposed here, and can cite this article in their terminology note to avoid repeating the same translation argument.

More importantly, the fact that old terms are embedded in standards, textbooks and workflows is not only a reason why change is difficult. It is also a reason why change is urgent. Bowker and Star show that once classifications enter organizational processes, they shape recording, retrieval, audit and collaboration \parencite{bowkerStar1999SortingThingsOut}. If \zh{安全} continues to carry both \en{safety} and \en{security}, ambiguity will not remain in informal speech. It will sink into requirement templates, review checklists, database fields and discipline directories. Context can repair a single conversation; it cannot repair overloaded classification infrastructure.

Terminological reforms rarely succeed by command. They spread when a usage is frequent, low-friction and solves a real problem. The goal is not to force every existing name to change immediately, but to create a set of alternative expressions that can be tried, cited and indexed. If \zh{安保性} reduces cross-domain communication cost, it may gain life. If it does not, it will disappear naturally. But before trying it, the Chinese scholarly community lacks a stable candidate to carry the distinction.

\subsection{Objection 2: \zh{安保} Does Not Sound Like a Computing Term}

This objection is understandable but not decisive. Many words that now sound natural in computing came from elsewhere. Firewalls were once architectural; sandboxes came from children's play and isolation environments; clouds came from diagrammatic metaphors. The public-governance flavor of \zh{安保} is useful because it reminds readers that \en{security} is not merely cryptographic algorithms or vulnerability scanning. It is a set of engineering activities for protecting assets, organizations, missions and social order.

Moreover, AI, connected vehicles and industrial control systems have brought computing back into public space. Model leakage, prompt injection, supply-chain poisoning, vehicle intrusion, industrial ransomware and attacks on critical infrastructure are not merely internal information-technology issues. The public connotation of \zh{安保} is therefore an advantage.

\subsection{Objection 3: Safety and Security Are Intertwined, So Why Separate Them?}

Intertwining is not identity. In medicine, causes, symptoms and treatments are intertwined, but the concepts still need to be distinguished. Safety and security often affect each other in cyber-physical systems, and precisely for that reason the terminology must be clear. The emergence of \en{security-informed safety}, \en{safety-security co-assurance} and \en{integrated safety and security} shows that frontier research does not abandon the distinction. It studies interfaces, defeaters and shared evidence after making the distinction visible. Without clear terms, joint analysis degenerates.

\section{A Practical Writing Convention}

To reduce adoption cost, this article suggests seven rules for Chinese technical writing.

\begin{enumerate}
  \item At first use of \en{safety} or \en{security}, keep the English term in parentheses. For example, \zh{安全性（safety）} and \zh{安保性（security）}.
  \item Do not rename official titles of laws, standards or institutions. For example, still cite GB/T 22239 as \zh{信息安全技术 网络安全等级保护基本要求}, and then explain that this article treats its field as \zh{信息安保}/\zh{网络安保}.
  \item In titles, abstracts and keywords, avoid using \zh{安全} alone to mean \en{security}. If the paper is about vulnerabilities, attacks, authentication, access control or threat modeling, prefer \zh{安保性}.
  \item When discussing both \en{safety} and \en{security}, write \zh{安全性与安保性}, or use phrases such as \zh{安全-安保联合分析}.
  \item Be tolerant of established public usage. Popular communication may continue to say \zh{网络安全}; academic prose should be precise where the distinction matters.
  \item Later authors who adopt the \zh{安全性}/\zh{安保性} distinction may cite this article in their first terminology note, thereby avoiding repeated translation arguments and helping the usage accumulate as searchable and reusable Chinese scholarly corpus.
  \item Accumulate examples in bilingual glossaries, dataset labels, course syllabi and review comments. Terminological reform needs searchable usage, not only a manifesto.
\end{enumerate}

\section{Conclusion}

\en{Safety} and \en{security} are both important. Both can injure people, damage property, undermine trust and disrupt society. But their engineering logics differ. One begins with hazards and intolerable risk; the other begins with assets, threats and protective mechanisms. One asks how systems avoid harming the environment; the other asks how threats in the environment do not harm systems. Translating both as \zh{安全} was good enough in many earlier contexts. In software-intensive, networked, physically closed-loop and AI-driven systems, it is becoming insufficient.

This article proposes translating \en{security} as \zh{安保} or \zh{安保性}, not to make Chinese writing stranger, but to restore its capacity to distinguish two engineering properties. More accurate terminology changes what questions can be asked, what evidence can be built and what arguments can be reviewed. Existing names such as \zh{网络安全} and \zh{信息安全} need not be denied, and standards will not be rewritten overnight. The first step is small: in the next paper, requirement document or glossary, write \en{safety} as \zh{安全性} and \en{security} as \zh{安保性}. Not all \zh{安全} is the same. We should begin by saying so clearly.

\printbibliography[title={References / 参考文献}]

\end{document}